\documentclass[sigconf]{acmart}

\usepackage{booktabs} % For formal tables
\usepackage{xspace}
\usepackage{soul}
\usepackage{balance}

\newcommand{\code}[1]{\texttt{#1}}

\newcommand{\helix}{{\sc Helix}\xspace}

\newcommand{\name}{\helix}

\newcommand{\dataprep}{data pre-processing\xspace}

\setcopyright{none}

\newenvironment{denselist}{
    \begin{list}{\tiny{$\bullet$}}%
    {\setlength{\itemsep}{0ex} \setlength{\topsep}{0ex}
    \setlength{\parsep}{0pt} \setlength{\itemindent}{0pt}
    \setlength{\leftmargin}{1.5em}
    \setlength{\partopsep}{0pt}}}%
    {\end{list}}

\newcommand{\topic}[1]{\vspace{2pt} \noindent{\bf #1.}}

\newcommand{\hidden}[1]{}

\DeclareMathOperator*{\argmin}{argmin}

\renewcommand\footnotetextcopyrightpermission[1]{}
\begin{document}
\title{Accelerating Human-in-the-loop Machine Learning: \\
Challenges and Opportunities}

\author{Doris Xin, Litian Ma, Jialin Liu, Stephen Macke, Shuchen Song, Aditya Parameswaran}
\affiliation{%
  \institution{University of Illinois,  Urbana-Champaign (UIUC)}
}
\email{{dorx0, litianm2, jialin2, smacke, ssong18, adityagp}@illinois.edu}

%%%%%%%%%      End of Authors       %%%%%%%%%%%%%

\renewcommand{\shortauthors}{D. Xin et al.}

% force figs to be on different pages
\renewcommand{\floatpagefraction}{0.1}

\begin{abstract}
%!TEX root=deem.tex

Development of machine learning (ML) workflows is a 
tedious process of {\em iterative experimentation}: 
developers repeatedly make changes to workflows until
the desired accuracy is attained.   
We describe our vision for a ``human-in-the-loop'' ML system that accelerates
this process:  
by intelligently tracking changes and intermediate
results over time, such a system can enable rapid iteration,
quick responsive feedback, 
introspection and debugging, and 
background execution and automation. 
We finally describe \helix, our preliminary attempt at such a 
system that has already led
to speedups of upto 10$\times$ on typical iterative workflows against
competing systems.

\end{abstract}

\maketitle

\section{Introduction}
\label{sec:intro}
%!TEX root=deem.tex

\begin{figure}[h]
\vspace{-10pt}
\centering
\includegraphics[width=0.4\textwidth]{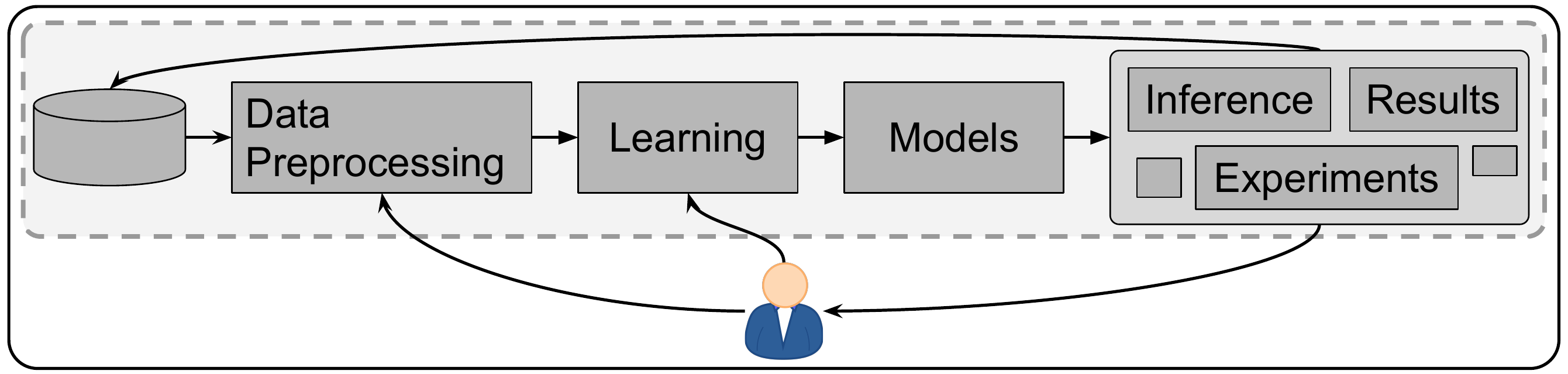}
\vspace{-10pt}
\caption{Development Cycle.}\label{fig:devCycle}
\vspace{-15pt}
\end{figure}

Due to the unpredictable nature of 
machine learning (ML) model performance,
developing ML applications 
relies on numerous iterations of trial-and-error---a {\em step-by-step process of experimentation}\footnote{www.nytimes.com/2014/08/18/technology/ for-big-data-scientists-hurdle-to-insights-is-janitor-work.html}.
The development process often begins with 
an initial workflow containing simple \dataprep and modeling steps.
Then, based on analysis of the resulting model, 
the developer modifies the workflow to improve performance. 
Examples of such modifications include adding/removing features, 
introducing new data sources,
switching from logistic regression to deep neural nets,
adding regularization to the model,
and changing the evaluation metrics.
Many such iterations take place 
between the conception and the deployment of the ML model,
with the developer as an integral component.
We model this process in Figure~\ref{fig:devCycle}, with
the dotted box representing one instance of the ML workflow,
with the developer ``in-the-loop'' using the end results as cues
for modifications. 

Unfortunately, most work on ML systems
has focused on specification and acceleration of the 
one instance of a given ML workflow 
(i.e., the dotted box in Figure~\ref{fig:devCycle}), 
without reasoning about the 
iterative ``human-in-the-loop'' aspect. 
By doing so, such systems have a number of deficiencies:
\begin{denselist}
\item {\bf \em Iterative reuse:} {\em Developers wait for minutes to 
hours even on small 
changes to the workflow.} 
On iterative changes (e.g., changing a feature
or regularization), developers rerun workflows
from scratch, since it is too cumbersome to identify and
reuse intermediate results.
\item {\bf \em Introspection between iterations:} \textit{Developers are not able to explain, debug, or understand performance.} 
Developers do not understand the impact
of changes they made to accuracy, and are not able to 
debug performance across the ML workflow. 
\item {\bf \em Leveraging think-time between iterations:} \textit{ML systems do not utilize think-time for background processing.}
Traditional ML systems do not take advantage of coding or thinking time
for background processing, such as trying out
workflow alternatives (e.g., changing regularization,
training, or doing automated feature selection).
\item {\bf \em Reducing iterative feedback latency:} \textit{ML systems do not optimize for quick feedback to developers.} 
Traditional ML systems do not provide rapid, approximate feedback that allows developers to make decisions before the computation is carried to full completion.
\item {\bf \em Automated cues for iterative changes:} \textit{ML systems do not provide novice developers
cues for subsequent changes.}
Traditional ML systems do not provide suggestions for next steps in the iterative
evolution of the workflow.
\end{denselist}
In this paper, we describe our vision for
an {\em accelerated human-in-the-loop machine learning (HILML) system}
that is targeted at eliminating the deficiencies identified above.
The ultimate goal of a HILML system is to {\em shorten the time
to obtain deployable models from scratch}. 
In addition, since we are focusing on the humans-in-the-loop,
we also aim to support declarative specification of ML workflows,
targeting novice users of ML, while also providing benefits
to expert users. 

We also present \name, our first attempt at  
a declarative, general-purpose HILML system 
aimed at accelerating iterative ML application development.
We briefly describe the programming interface and system architecture,
and demonstrate the effectiveness of \name 
with preliminary results on multiple applications.

\topic{Related Work}
Prior work has recognized the importance of studying many aspects of HILML, 
such as iterative model building~\cite{vartak2015supporting}, 
model management both in general~\cite{vartak2016m} 
and specifically tailored for deep learning~\cite{miao2017towards}, 
dataset versioning~\cite{van2017versioning},
and model sharing~\cite{miao2017model}. 
We complement existing work with a fresh set of new research challenges.
In the general space of ML systems, 
a number of recent work focus on using declarative programming to improve usability~\cite{zhang2015deepdive,kraska2013mlbase,sparks2017keystoneml}.
While some systems aim to support end-to-end ML in a general setting~\cite{pedregosa2011scikit, sparks2017keystoneml, meng2016mllib, kraska2013mlbase},
many focus on a special component of the process, 
such as model selection~\cite{sparks2015tupaq},
feature selection~\cite{Zhang2016Columbus},
and feature engineering~\cite{zhang2015deepdive}.
Unlike \name, these systems are focused on optimization in the single-execution setting,
neglecting to consider challenges and opportunities associated with iterative development.

\section{Optimizing Human-in-the-loop ML}
\label{sec:opt}
%!TEX root=deem.tex

In this section, we first
describe a concrete, unifying model 
to ground our discussion and describe prerequisites
that any HILML system must obey.
Then, in the next subsection, we
describe concrete research directions
that are enabled by this unifying model.

\subsection{Prerequisites: Usability and Model}\label{sec:prereq}

\topic{Usability: Declarativity and Generalization} A basic requirement for HILML systems is that
it can generalize across use-cases (spanning
applications in the social and natural sciences, for example), workflow design decisions (from supervised to unsupervised learning,
and from linear regression to deep neural nets, for example), 
and expertise levels (from novice to expert).
The latter concern is especially important
given the demand for ML and AI in a host of new emerging
data-driven disciplines. 
To do so, such a system must accept a declarative or semi-declarative
specification (such as recent tools~\cite{sparks2017keystoneml,zhang2015deepdive,ghoting2011systemml}), while also be able to embed
arbitrary imperative code. 

\topic{Workflow DAG Model} Given this declarative specification (with black-box
UDFs), a HILML system must be able to capture and abstract 
a specific instance of a 
workflow as a {\em DAG of intermediate data items}, where
the nodes in the DAG correspond to the output of specific
operators,
and the edges indicate input-output relationships 
between operators. 
Via the declarative specification, the HILML system can identify 
the logical operator corresponding to each node
in the workflow (e.g., data preparation or model training), 
giving the system
a comprehensive understanding of the workflow. 

Between two iterations of the workflow, a specific intermediate
data item is deemed to be identical in both iterations
if the source code for the corresponding operator
has not changed, and recursively the 
parents of that data item are identical.
A source data item is deemed to be identical if the 
underlying data has not changed between iterations.

To be able to tell what has changed or what has not changed
between iterations is not straightforward,
however we can detect file-level changes to the source data items,
and we can detect code-level changes to the declarative specification;
changes to external libraries may be harder to detect, but are likely
to not be so frequent.

Our workflow model provides us a valuable starting point for 
a HILML system---by detecting what is same and what is different
across iterations allows a HILML system to understand
how the workflow has evolved over iterations.

\subsection{HILML Research Challenges}\label{sec:challenges}

We now describe a number of research challenges
that are enabled by our workflow model. 

\topic{Intermediate Results Reuse}
To enable effective reuse of intermediate data items 
across iterations,
we must be able to answer two questions:
\begin{denselist}
\item[1)] What intermediates should be materialized in the current execution to speed up future iterations through reuse?
\item[2)] What intermediates should be reused to minimize run time given materialized intermediates?
\end{denselist}
The answers to both questions can be represented as 
a subset of nodes in a specific instance of the workflow DAG.
For 1, notice first that materializing all intermediates
may not be beneficial due to high cost that may 
outweigh any potential benefits.
In fact, the answer to 1 is contingent on 
a number of prediction problems.
Since not all intermediates will be 
reusable given the dynamic nature of the workflow, 
modeling future savings 
incurred by reuse requires predicting iterations,
both in terms of the number of 
iterations and the specific modifications.
For example, materializing an intermediate data item 
below a workflow 
portion that the developer is actively modifying 
is not likely to be useful, since it may be
rendered redundant by the changes to the workflow.
Predicting which portion of the workflow a developer
may modify next (e.g., maybe the developer is done with feature
engineering and have moved onto model tuning) can also
help with this decision.

Comparatively, the answer to 2 is more 
self-contained due to the lack of uncertainty.
The complexity for 2 lies within the fact 
that it needs to take place during compile time, 
so the system needs to estimate the run time 
and output size of the operators.

\topic{Introspection: analyzing the impact of changes}
Knowing statistics 
such as how certain changes to the workflow 
have impacted  
prediction accuracy and overall run time 
helps developers maximize the utility of future iterations.
This goes beyond tracking metrics and data 
associated with each model version 
and delves into exploring the causal relationship 
between performance and specific changes.
Recognizing such relationships 
requires semantic understanding of the workflow,
which enables \textit{logical comparison} of the different 
workflow versions
(e.g., version 2 adds a feature and regularization to version 1).
The workflow representation provides a means 
for visualizing the logical difference 
between two versions of the workflow.

Furthermore, we can help developers 
identify the set of workflow versions 
that are the most pertinent to a 
specific performance gap.
This can be framed as a path finding problem 
in the space of workflow versions,
with distance between versions reflecting 
the amount of logical changes.
Specific challenges include distance metric 
design and characterization of the desired paths.

\topic{Automated background search during think time}
Based on past usage and semantic understanding of the workflow,
the system should be able to 
automatically identify modifications to the current workflow
that lead to potential improvements on the metrics of interest.
These changes can be tested in the background 
while the system is idle (e.g., the developer is writing code or thinking). 
Note that this ranges beyond grid search 
for model hyperparameters~\cite{kraska2013mlbase}
to include modifications to the \dataprep 
components of the workflow as well.
The goal is not to cut the human out of the loop, 
but rather to remove tedious, mechanical iterations 
from active development time.
The developer should spend their time on 
applying domain knowledge to improve the application 
instead of exhaustively testing out known tricks.

\topic{Quick feedback: end-to-end optimization}
Compared to intra-operator optimization 
(e.g., speeding up model training),
end-to-end workflow optimization 
has a much greater potential for hastening iterations
since it is able to capture the higher order, inter-operator inefficiencies 
missed by intra-operator optimization.
Optimizing workflows end-to-end is challenging 
due to the difficulty of analyzing relationships between operators.
Sometimes operators within a workflow can be written 
in different languages or using different libraries.
To enable general end-to-end optimization,
we need a common, framework-independent abstraction of operators 
capable of modeling how data logically flows between operators.
Here is a concrete example of end-to-end ML workflow optimization: 
if a model is sparse (many zero weights), 
being able to identify the operators corresponding to zero-weight features 
allows us to prune a large portion of the workflow 
without compromising accuracy.

\topic{Quick feedback: approximate workflow execution}
A developer's decision to keep changes made in an iteration 
depends on the \textit{relative} performance to the previous iteration's. 
Thus, the results obtained in each iteration do no need to be precise, 
as long as they accurately indicate the performance trend.
%This property presents an opportunity for approximate computing to accelerate iteration.
The system should provide mechanisms 
to allow developers to trade \textit{redundant precision} for speed,
especially when the data size is large. 
Approximate computing techniques 
such as sampling and using low precision floats 
can be applied to achieved the desired tradeoff 
between precision loss and speedup.
This allows developers to test out the same 
number of changes in a fraction of the time,
with little compromise to their ability to accurately judge the effect of each change.

\topic{Automated cues for novice: change recommendations}
While expert ML developers may have good intuition how to improve the workflow,
inexperienced users can greatly benefit from suggestions on what to try next.
Concretely, the system should suggest to the developer 
what operators to add/delete/modify and the expected outcome.
The iteration prediction model discussed above can be adapted for this purpose.
Developing such a model requires gathering data on 
how developers iterate on workflows in various domains,
which can be a difficult task since publications tend to focus on results instead of the process.
Ideally, the model should consider workflow and data characteristics 
as well as user skills and system settings.
Furthermore, it should learn from the results of previous iterations.
Note that we need to be careful 
not to trap the developer in a local optimum 
from this feedback loop.

\section{\name for Human-in-the-loop ML}
\label{sec:overview}
%!TEX root=deem.tex

\name is our first attempt at a HILML system,
satisfying the prerequisites outlined in Section~\ref{sec:prereq},
and partially addressing the first research challenge in Section~\ref{sec:challenges}.
\name has a declarative programming interface that is concise yet expressive;
it uses the workflow DAG model 
to enable both end-to-end and 
cross-iteration optimizations.
In this section we provide a brief overview of the system architecture,
followed by preliminary results on performance compared to related state-of-the-art systems.

\subsection{System Architecture}

\begin{figure}[h]
\centering
\includegraphics[width=0.37\textwidth]{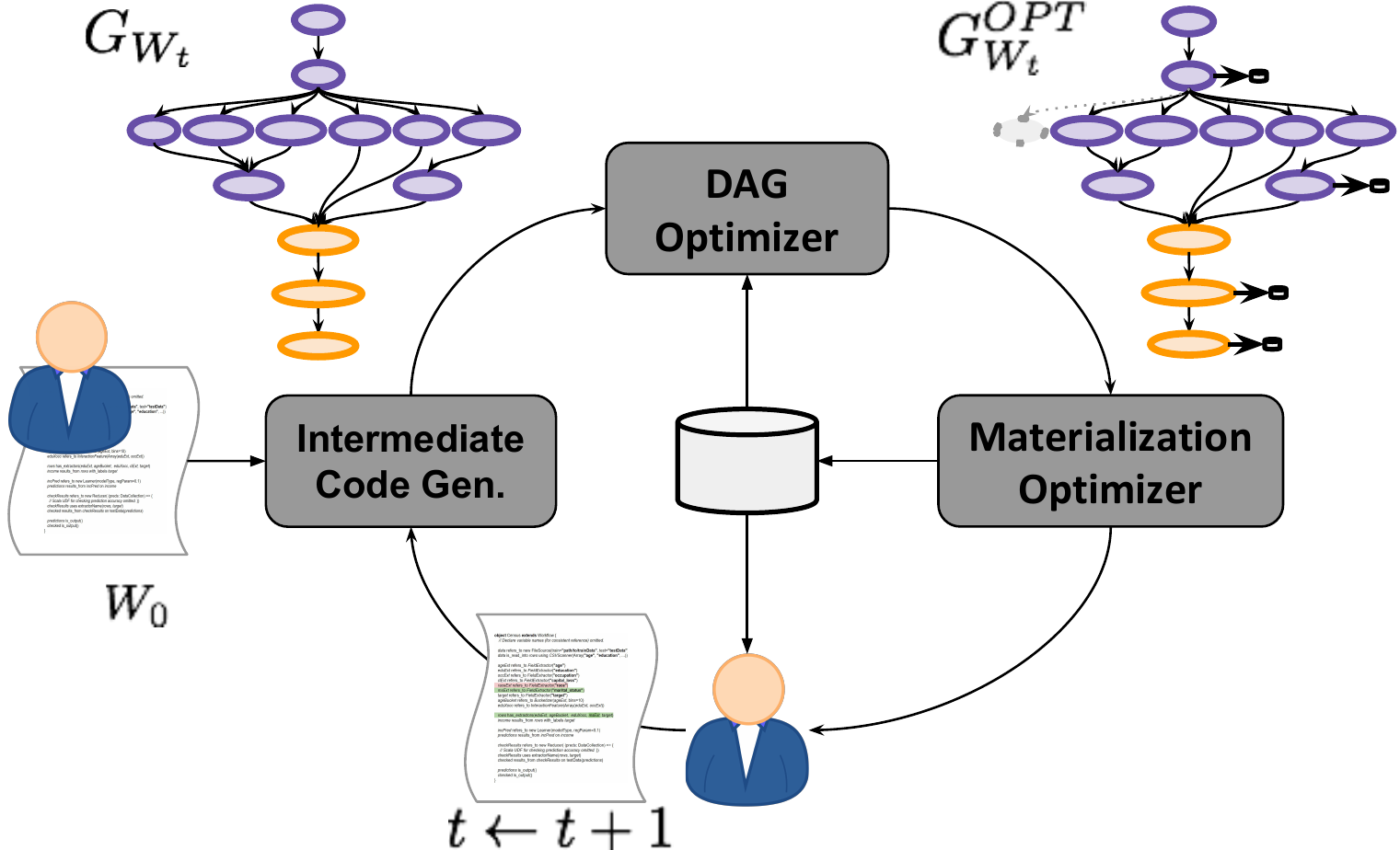}
\vspace{-10pt}
\caption{Lifecycle of a Workflow in \name.}\label{fig:hcycle}
\vspace{-10pt}
\end{figure}

\name consists of a programming interface, 
a compiler for the client application code, 
and an execution engine. 
The overarching optimization objective is 
to minimize end-to-end execution time across iterations
through intermediate result reuse 
and redundant operator pruning.
As shown in Figure~\ref{fig:hcycle}, 
the \textit{materialization optimizer} 
decides what intermediate results to persist to disk 
at run time to accelerate subsequent iterations, 
while the \textit{DAG optimizer} decides what set of intermediates to reload from disk 
and what operators to prune 
at compile time to speed up the current iteration.
We describe the programming interface and the optimization problems below.

\subsubsection{Programming Interface}
\begin{figure}[h]
\centering
\vspace{-10pt}
\includegraphics[width=0.44\textwidth]{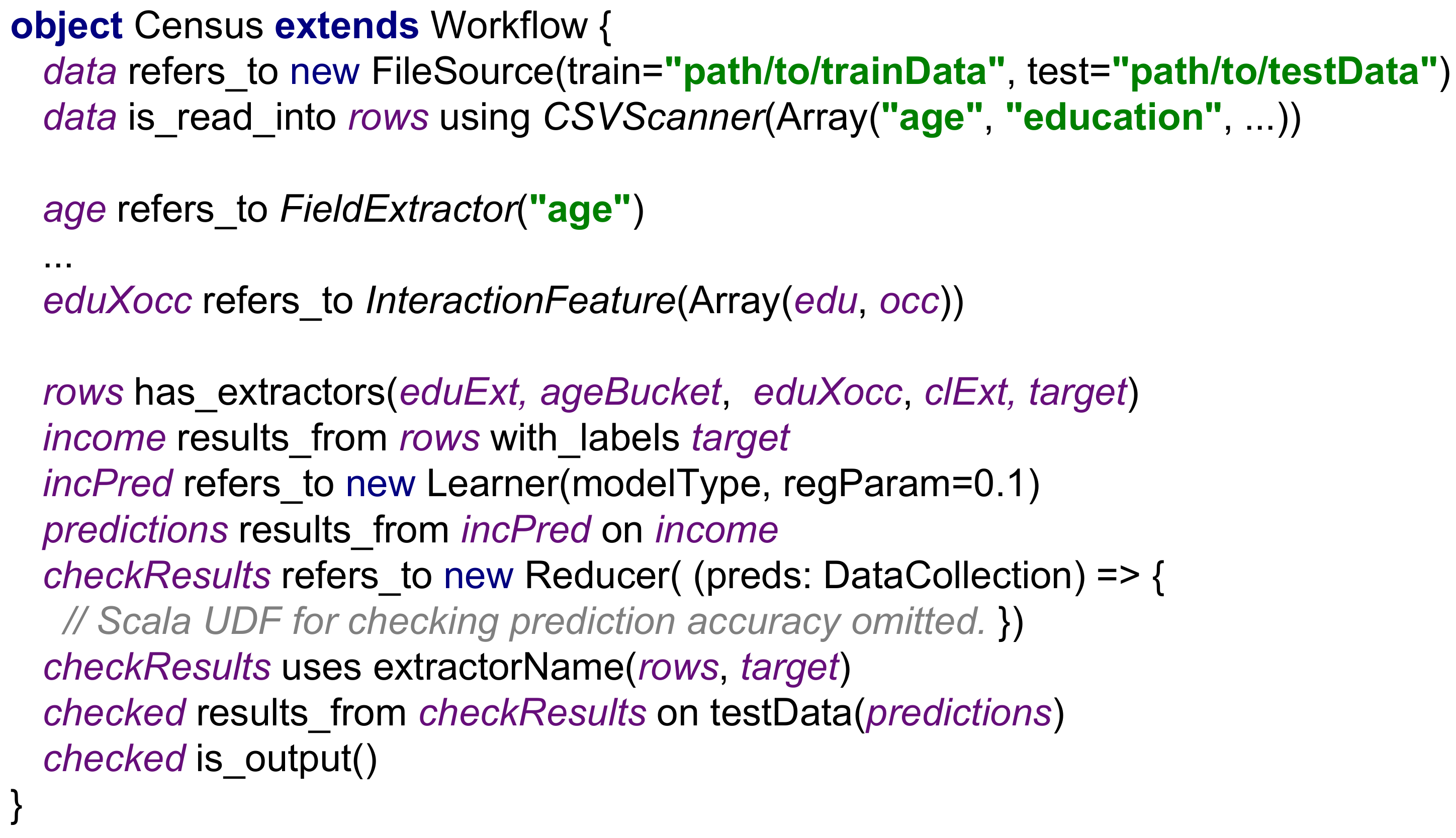}
\vspace{-10pt}
\caption{Sample code in \name DSL}\label{fig:code}
\vspace{-18pt}
\end{figure}

\name's programming interface is a DSL embedded in Scala,
where the statements are declarative 
and designed to mimic natural language with infix expressions.
A single Scala interface named \code{Workflow} is used to program the entire workflow
% Operators
\name provides a handful of composable and extensible operator types 
to handle variability and complexity in both \dataprep and the learning task,
ranging from fine-grained to whole-dataset feature engineering, 
supervised learning to information extraction.
% imperative code integration
Imperative Scala code can be directly embedded into the DSL 
for user-defined functions (UDFs),
similar to inline SQL UDF registration in SparkSQL~\cite{armbrust2015sparksql}.
% Data structures
\name uses two distinct data structures to handle \dataprep and ML, 
so that data is kept in human-readable formats for easy feature engineering 
and automatically transformed into ML compatible formats for learning.
% Outcome
We have used the DSL to successfully implement workflows 
in {\em social sciences, information extraction, computer vision, and natural sciences,}
spanning a wide array of \dataprep and ML use cases. 
Figure~\ref{fig:code} shows a code snippet of the DSL used to program an income prediction task.

\subsubsection{Compilation}
\name compiles each \code{Workflow} into a DAG of operators.
Using both the compiled DAG and relevant data from disk, 
the DAG optimizer performs three tasks:

\topic{Detect changes}
\name automatically detects the set of operators 
that have changed since the last iteration  
and marks them for mandatory recomputation.
All operators are named, 
and the DAG optimizer performs light program analysis 
to compare operators with the same names 
across iterations for change detection.

\topic{Prune redundant operators}
\name prunes extraneous operators by 
applying dataflow analysis to identify 
operators that do not contribute to the final output. 
This feature alleviates the burden of 
manually removing dead code 
to avoid redundant computation 
when data dependencies change.

\topic{Compute the optimal reuse policies}
Loading all reusable intermediates from disk 
is not always the optimal decision 
for minimizing overall run time.
For example, if an operator has a large output but a short compute time,
then it is more time-efficient to load its input and recompute.
Loading the results of an operator does, however,
allow us to prune its inputs,
which could have a cascading effect leading to large savings.
We can formally model this problem 
as assigning states to the nodes in the workflow DAG.
Each node can be assigned one of three states \{compute, load, prune\}, 
and the assignments must satisfy the pruning constraint that
a node in the \textit{compute} state 
must not have parents in the \textit{prune} state.
The objective is to find a legal state assignment $s^* = $
\vspace{-5pt}
\begin{equation}\label{eq:reuse}
\argmin\limits_{s} \sum\limits_{n_i \in N} \mathbb{I}\{s(n_i) = compute \}c_i + \mathbb{I}\{s(n_i) = load \} l_i
\vspace{-5pt}
\end{equation}
where $s(n_i)$ is the state of node $n_i$, 
and $c_i$ and $l_i$ are the compute and load time, respectively.
This problem cannot be solved using a single pass algorithm because of the pruning constraint. 
We devise an efficient PTIME algorithm to solve Eq (\ref{eq:reuse}) optimally 
by proving that it is polynomial time reducible to {\sc Max-Flow}~\cite{dorx2017}.

\subsubsection{Execution Engine}
%!TEX root=deem.tex

\name carries out the optimal physical plan produced by the DAG optimizer 
using Spark~\cite{zaharia2012resilient} for distributed data processing.
Note that the core algorithms and optimization techniques in \name 
are independent of the data processing platform.
Lightweight wrappers can be written 
to support other data processing frameworks such as Tensorflow~\cite{abadi2016tensorflow}. 

During execution, the \textit{materialization optimizer}, 
shown in Figure~\ref{fig:hcycle}, 
handles the optimization problem of 
choosing the intermediates to materialize for reuse in future iterations,
under a maximum storage constraint.
As discussed in Section~\ref{sec:opt}, 
the benefit of materializing an intermediate result 
is dependent on its likelihood of being reused in future iterations.
Even with the simplifying assumptions that 
1) there will be only one more iteration
and 2) all intermediate results are reusable in the next iteration,
the problem is still {\sc NP-Hard}, 
as we show through a reduction from {\sc Knapsack}~\cite{dorx2017}.

Another complicating factor is that
we must make the decision to materialize \textit{online},
i.e., immediately after an operator has completed execution, 
since deferred decisions are prohibitive as they require caching multiple intermediate results.
We propose a simple cost model 
to achieve an approximate solution 
while respecting the online constraint.
Given the load cost $l_i$ and compute cost $c_i$ for each operator $n_i$,
the cost $r_i$ of materializing $n_i$ is defined as
\vspace{-5pt}
\begin{equation}
 ( c_i + \sum\limits_{n_j \in A(n_i)} c_j ) - 2l_i
 \vspace{-5pt}
\end{equation}
We materialize $n_i$ if $r_i$ is negative and 
$l_i$  is less than the remaining storage.
The model naively assumes that loading $n_i$ 
prunes all of its ancestors from the DAG.
While this is untrue as discussed above, we cannot hope to do better given the online constraint.
This model has been effective in experimental studies, to be discussed next.

\begin{figure}[h!]
\centering
\vspace{-10pt}
\includegraphics[width=0.16\textwidth]{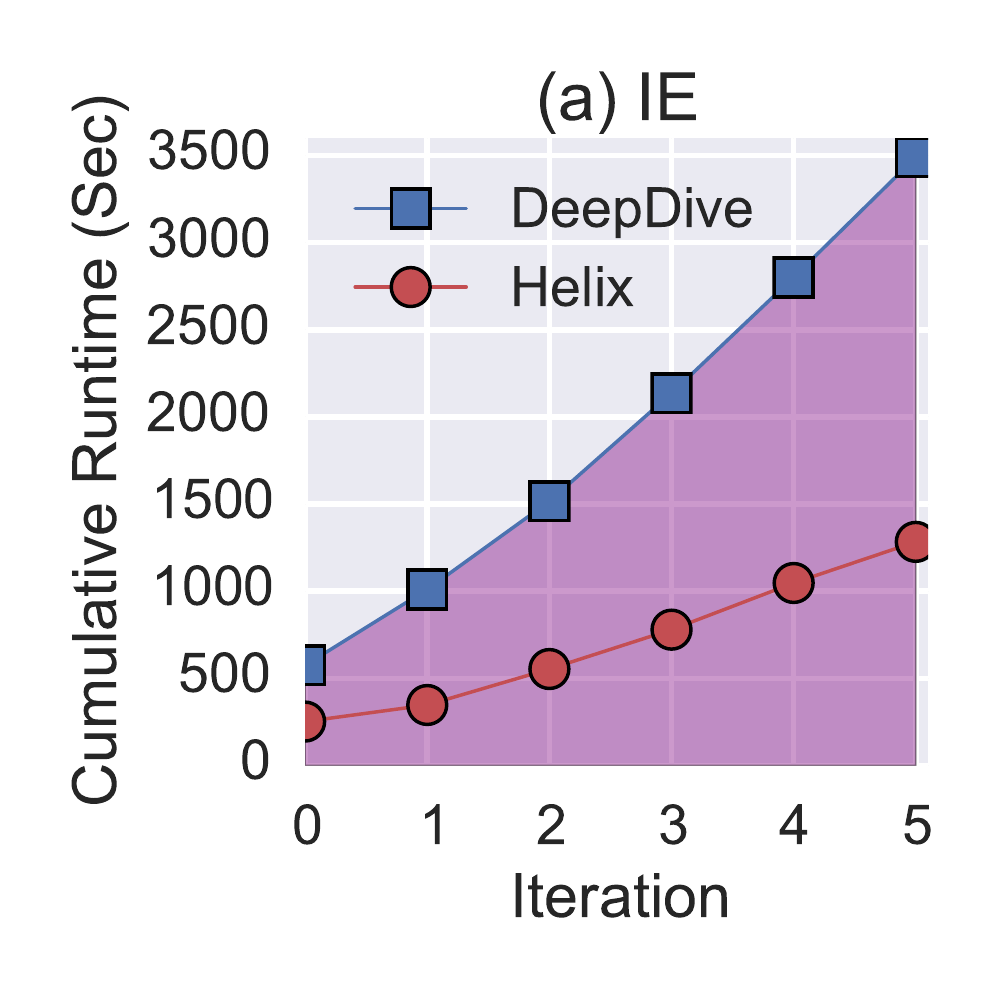}
\includegraphics[width=0.24\textwidth]{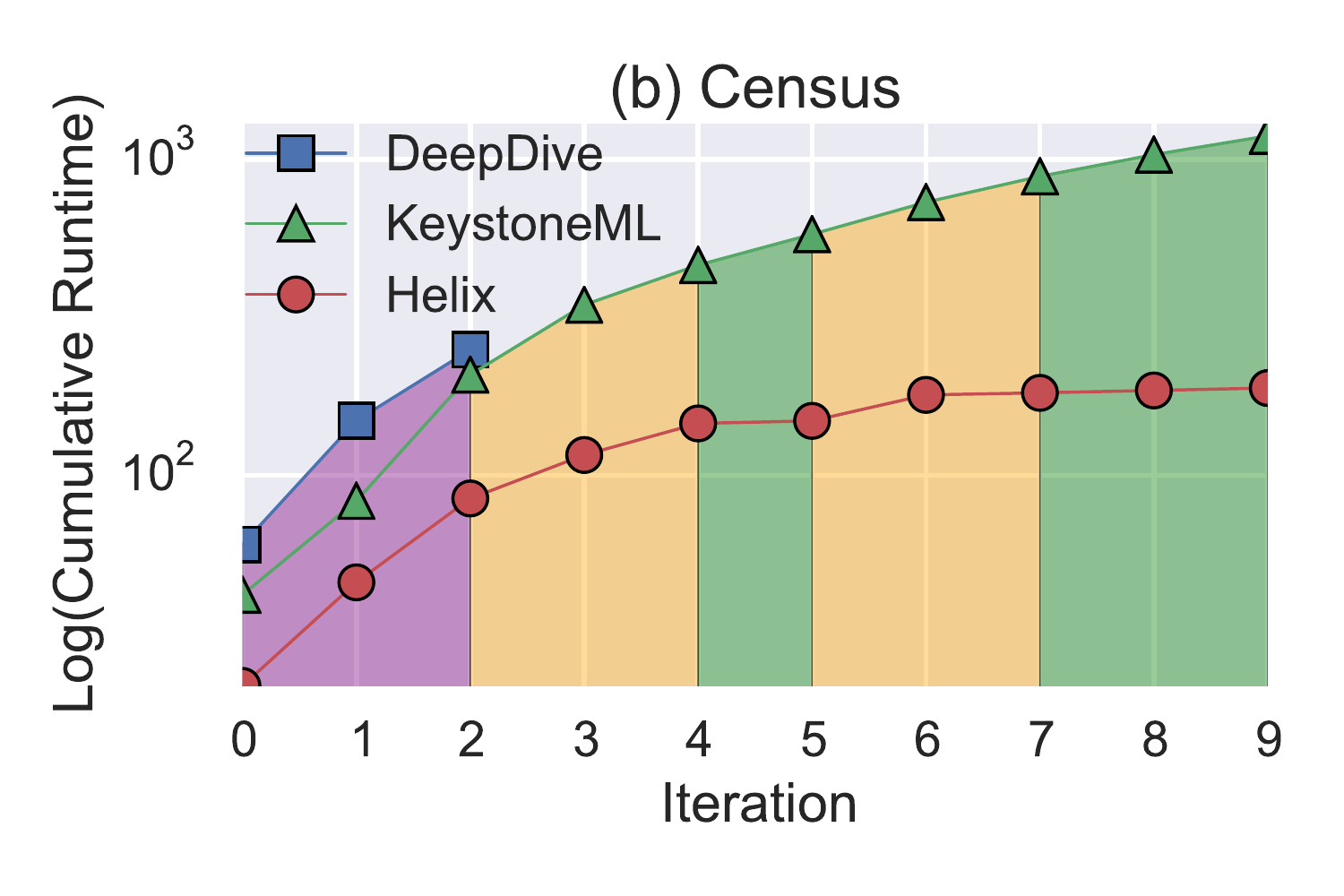}
\vspace{-18pt}
\caption{Logscale cumulative runtime comparison with (a) DeepDive on an IE task.
(b) DeepDive and KeystoneML on a classification task.}
\label{fig:perf}
\vspace{-18pt}
\end{figure}

\balance
\subsection{Performance Gains}
We present preliminary results
comparing  \name with two similar ML systems, DeepDive~\cite{zhang2015deepdive} and KeystoneML~\cite{sparks2017keystoneml},
on an application in information extraction (IE) and another on classification.
KeystoneML is not designed to handle IE tasks, 
hence absent in Figure~\ref{fig:perf}(a).
DeepDive has no data for iteration $>2$ in Figure~\ref{fig:perf}(b)
due to inconfigurable ML and evaluation components.
The color under the curve indicates the type of change in each iteration,
with purple for \dataprep, orange for ML, and green for evaluation. 
The frequencies of each iteration type is determined 
using statistics collected on 105 applied ML papers~\cite{dorx2017}.

In contrast to DeepDive's materialize-all approach, 
\name judiciously materializes only intermediates that help reduce future run time,
resulting in a {\em 60\% reduction} in cumulative run time,
as shown in Figure~\ref{fig:perf}(a).
On the classification task, \name achieves
{\em an order of magnitude reduction} in cumulative run time
compared to both KeystoneML, which materializes no intermediates, and DeepDive,
as shown in Figure~\ref{fig:perf}(b).
We see in both workflows that \dataprep iterations (purple) have the highest iteration run times,
while evaluation (green) has the lowest, 
proportional to the amount of mandatory recomputation for each iteration type.

\section{Conclusions}
\label{sec:conclusion}
We presented our vision for an efficient end-to-end ML system 
focused on supporting iterative, human-in-the-loop workflow development.
We identified specific research problems to accelerate and automate workflow development 
and introduced \name---our first attempt at addressing some of these problems.

\bibliographystyle{abbrv}
\bibliography{sigproc}

\begin{thebibliography}{10}

\bibitem{abadi2016tensorflow}
M.~Abadi et~al.
\newblock Tensorflow: A system for large-scale machine learning.
\newblock In {\em OSDI}, volume~16, pages 265--283, 2016.

\bibitem{armbrust2015sparksql}
M.~Armbrust et~al.
\newblock Spark sql: Relational data processing in spark.
\newblock In {\em SIGMOD}, 2015.

\bibitem{ghoting2011systemml}
A.~Ghoting et~al.
\newblock Systemml: Declarative machine learning on mapreduce.
\newblock In {\em ICDE}, 2011.

\bibitem{kraska2013mlbase}
T.~Kraska et~al.
\newblock Mlbase: A distributed machine-learning system.
\newblock In {\em CIDR}, 2013.

\bibitem{meng2016mllib}
X.~Meng et~al.
\newblock Mllib: Machine learning in apache spark.
\newblock {\em JMLR}, 2016.

\bibitem{miao2017model}
H.~Miao et~al.
\newblock On model discovery for hosted data science projects.
\newblock In {\em DEEM}, 2017.

\bibitem{miao2017towards}
H.~Miao et~al.
\newblock Towards unified data and lifecycle management for deep learning.
\newblock In {\em ICDE}, pages 571--582. IEEE, 2017.

\bibitem{pedregosa2011scikit}
F.~Pedregosa et~al.
\newblock Scikit-learn: Machine learning in python.
\newblock {\em JMLR}, 2011.

\bibitem{sparks2015tupaq}
E.~R. Sparks et~al.
\newblock Tupaq: An efficient planner for large-scale predictive analytic
  queries.
\newblock {\em arXiv preprint arXiv:1502.00068}, 2015.

\bibitem{sparks2017keystoneml}
E.~R. Sparks et~al.
\newblock Keystoneml: Optimizing pipelines for large-scale advanced analytics.
\newblock In {\em ICDE}, 2017.

\bibitem{van2017versioning}
T.~van~der Weide et~al.
\newblock Versioning for end-to-end machine learning pipelines.
\newblock In {\em DEEM}, 2017.

\bibitem{vartak2015supporting}
M.~Vartak et~al.
\newblock Supporting fast iteration in model building.
\newblock In {\em NIPS Workshop LearningSys}, 2015.

\bibitem{vartak2016m}
M.~Vartak et~al.
\newblock Modeldb: a system for machine learning model management.
\newblock In {\em HILDA}, page~14. ACM, 2016.

\bibitem{dorx2017}
D.~Xin et~al.
\newblock Helix: Holistic optimization for accelerating iterative machine
  learning.
\newblock {\em Technical Report
  http://data-people.cs.illinois.edu/helix-tr.pdf}, 2018.

\bibitem{zaharia2012resilient}
M.~Zaharia et~al.
\newblock Resilient distributed datasets: A fault-tolerant abstraction for
  in-memory cluster computing.
\newblock In {\em NSDI}, 2012.

\bibitem{zhang2015deepdive}
C.~Zhang.
\newblock {\em DeepDive: a data management system for automatic knowledge base
  construction}.
\newblock PhD thesis, The University of Wisconsin-Madison, 2015.

\bibitem{Zhang2016Columbus}
C.~Zhang, A.~Kumar, and C.~R{\'e}.
\newblock Materialization optimizations for feature selection workloads.
\newblock {\em ACM Trans. Database Syst.}, 2016.

\end{thebibliography}

\end{document}